\pdfoutput=1
\documentclass[a4paper,10pt]{article}
\usepackage[utf8]{inputenc}
\usepackage{graphicx}
\usepackage{mathptmx}      
\usepackage{natbib}
\usepackage[colorlinks,citecolor=blue,urlcolor=red]{hyperref}
\usepackage{amsmath, bm, amssymb}
\usepackage{mathtools}
\usepackage{courier}
\usepackage{caption}
\usepackage{subcaption}
\usepackage{authblk}

\title{Prediction of tool-wear in turning of medical grade cobalt chromium molybdenum alloy (ASTM F75) using non-parametric Bayesian models}
\author[1]{Damien McParland}
\author[2]{Szymon Baron}
\author[3]{Sarah O'Rourke}
\author[2]{Denis Dowling}
\author[2]{Eamonn Ahearne}
\author[1,4]{Andrew Parnell}

\affil[1]{School of Mathematics and Statistics, University College Dublin, Ireland.}
\affil[2]{School of Mechanical and Materials Engineering, University College Dublin, Ireland.}
\affil[3]{Central Statistics Office of Ireland, Dublin, Ireland.}
\affil[4]{Insight Centre for Data Analytics, University College Dublin, Ireland.}

\date{}
\begin{document}

\maketitle

\begin{abstract}
We present a novel approach to estimating the effect of control parameters on tool wear rates and related changes in the three force components in turning of medical grade Co-Cr-Mo (ASTM F75) alloy. Co-Cr-Mo is known to be a difficult to cut material which, due to a combination of mechanical and physical properties, is used for the critical structural components of implantable medical prosthetics. We run a designed experiment which enables us to estimate tool wear from feed rate and cutting speed, and constrain them using a Bayesian hierarchical Gaussian Process model which enables prediction of tool wear rates for untried experimental settings. The predicted tool wear rates are non-linear and, using our models, we can identify experimental settings which optimise the life of the tool. This approach has potential in the future for realtime application of data analytics to machining processes. 
\end{abstract}

 \section{Introduction}
    \label{sec:intro}

	Technological advances in high-end manufacturing and ever stringent international standards drive the growth in demand for complex components and high precision tolerances. Automotive, aeronautical and medical device industries often rely on a business platform that involves the use of CAD/CAM technologies and multi-axis computer numerically controlled machine tools. The overall capability of this process chain is often constrained by the core cutting process and the fundamental mechanisms that result in thermal error, deflection and tool-wear \citep{Altintas2012, Shaw2005}. The constraints imposed by the underlying fundamental mechanisms of cutting are further exacerbated when machining difficult to cut (DTC) materials such as cobalt, nickel or titanium based superalloys \citep{Williams2008}. 

        Co-Cr-Mo alloys are well established as materials for medical devices and orthopaedic implants in particular \citep{Williams2008, Aljabbari2014}. High mechanical strength and wear resistance make them suitable even for applications in metal-on-metal bearing components while biocompatibility allows for long-term incorporation in the human musculoskeletal system  \citep{Williams2008}.  In developed economies, factors such as increased physical activity, rising rates of obesity and increased life expectancy are considered as the main drivers of growth in demand for knee and hip arthroplasties \citep{Tozzi2015, CDC}. Given the risks associated with revision surgeries, the Co-Cr-Mo components are expected to perform to the required standard for up to 20 years \citep{Beuchel2001}. This imposes the need for high precision tolerances and surface finish while growth in demand presents challenges for high levels of process throughput.

        Tool-wear, defined as gradual loss of the tool material from the areas of contact with the workpiece, directly affects these process measures  \citep{Altintas2012, Shaw2005}. Despite the importance of the application, market size and future challenges, there are very few publications on machining of Co-Cr-Mo alloys. Furthermore any predictive models of tool-life are often unique for particular tool-material-cutting fluid-machine configuration  \citep{Shaw2005, Halpin2012}. More recently it has been advocated that the era of Industry 4.0 and/or digital manufacturing (DM) may contribute to the deconvolution of inherent complexity of processes \citep{MTTRF16}. This may be achieved through incorporation of disruptive technologies such as new smart sensors, the internet of things, advanced data analytics and cloud computing \citep{MTTRF16,Byrne20161}. The principle is to follow a systematic progression in `levels of intelligence' as shown on Figure~\ref{fig:Roadmap} below. Incorporation of disruptive technologies into machining processes requires input from sensing equipment, signal processing and decision making support systems.

    \begin{figure}[h]
      \centering
      \includegraphics[width=0.7\textwidth]{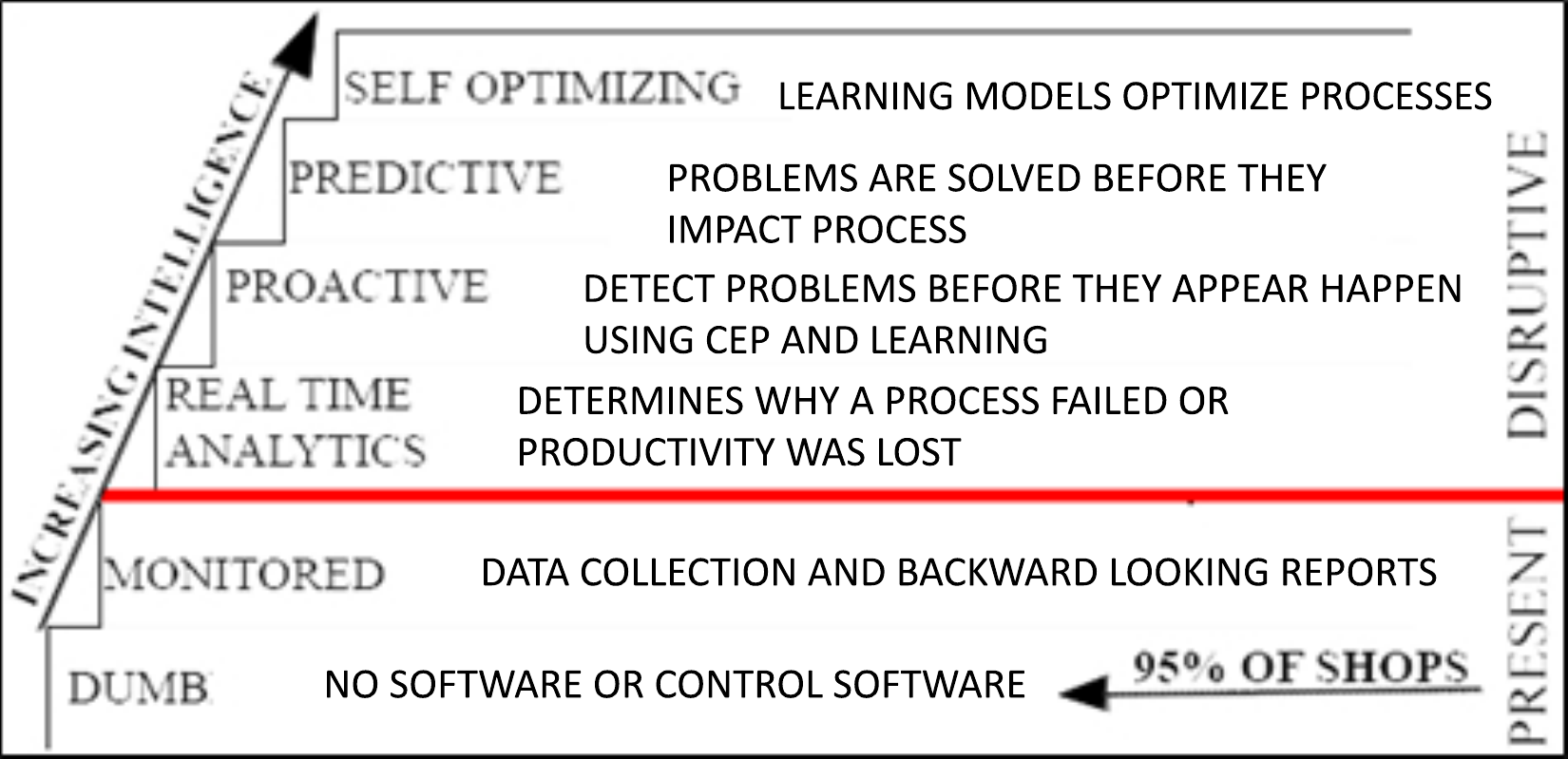}
      \caption{Roadmap for machining process control.}
      \label{fig:Roadmap}
    \end{figure}

	Our previous work detailed the properties of Co-Cr-Mo alloys and indicated that these alloys should be classified as DTC materials \citep{Baron2015c}. This paper now presents an approach that utilises non-parametric Bayesian hierarchical modelling to determine the rates of tool wear and estimate the cutting lengths at which changes  in these rates occur, such as the length of cut when tool failure occurs. For each set of control parameters, the progression of tool wear is modelled against the profile of the cutting forces recorded by a piezoelectric dynamometer. As part of our suite of output, we present response surface plots and estimated tool life values.

    The paper proceeds as follows. Section~\ref{sec:litreview} provides an overview of other research done in these areas. Section~\ref{sec:methods} cover the experimental design and statistical algorithms used to analyse the collected data are explained in detail. In Section~\ref{sec:results} we present the results of applying these methods to the Co-Cr-Mo ASTMF75 cutting data. We finish with a discussion in Section~\ref{sec:discussion}.

    \section{Previous Work}
\label{sec:litreview}

\begin{figure}[h]
      \centering
      \includegraphics[width=0.7\textwidth]{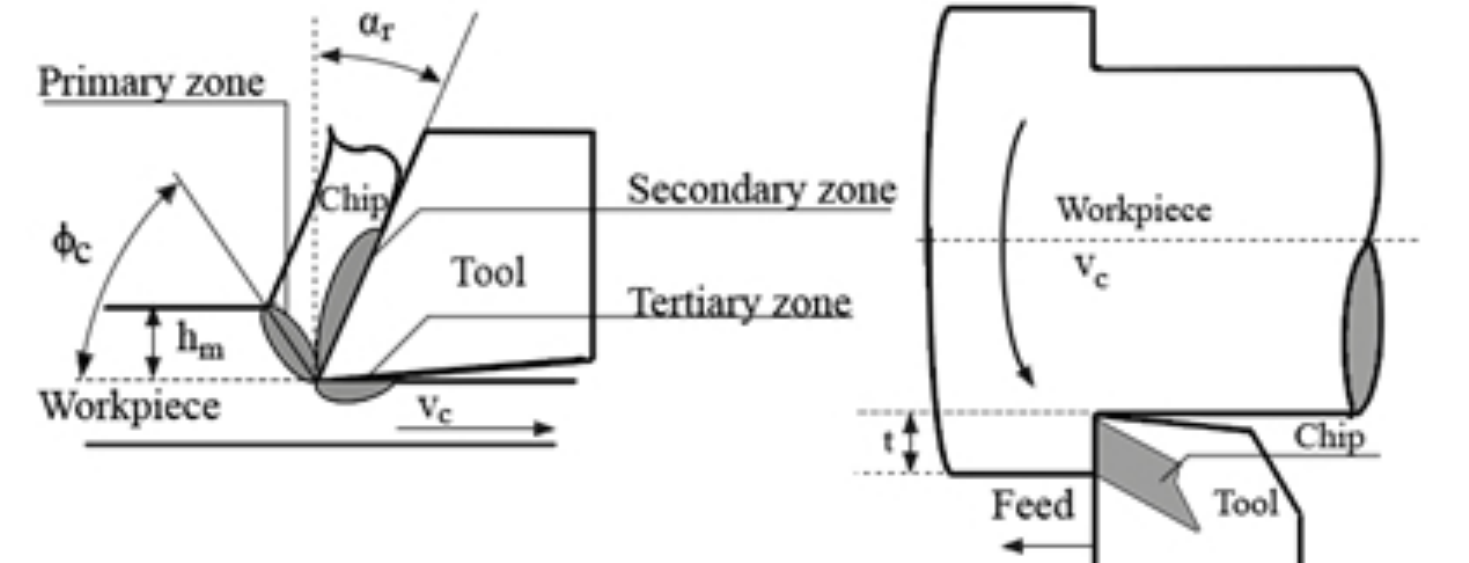}
      \caption{Schematic typical wear patterns}
      \label{fig:ChipFormation}
    \end{figure}

\cite{Shaw2005} defines metal cutting as the removal of a macroscopic chip by a tool with a defined cutting edge. Turning, milling and drilling are common cutting operations. In turning, ``a single point tool removes unwanted material to produce a surface of revolution'' \citep{Shaw2005}. The multi-physics mechanisms that take place during this process may be explained with reference to simplified orthogonal cutting as proposed by Merchant and enhanced by his model of the cutting mechanics \citep{Merchant1945, Markopoulos2013}. In the primary deformation zone as shown in Figure~\ref{fig:ChipFormation} the workpiece material is subject to extreme strain rates of $2.0\cdot 10^4s^{-1}$ and temperatures of 1200$^\circ$C  \citep{Jaspers2002} in machining of Inconel 718 \citep{Dudzinski2004}.
 
\begin{figure}[h]
      \centering
      \includegraphics[width=0.4\textwidth]{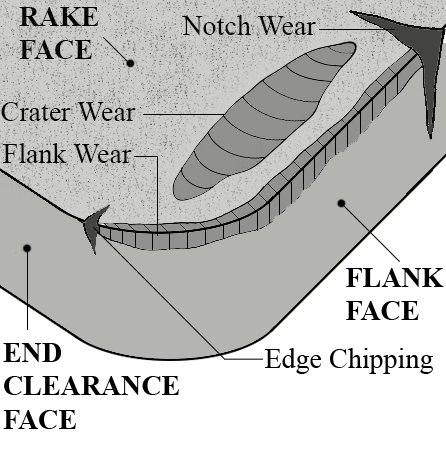}
      \caption{Schematic of 2D chip formation (left) and a single point turning (right).}
      \label{fig:Characteristic tool wear patterns}
    \end{figure}

The primary source of heat is the deformation in the primary shear zones while the secondary and tertiary sources of heat are primarily due to friction at (a) the tool-chip interface in the secondary shear zone and (b) the tool-workpiece interface at the tertiary shear zone \citep{Abukhshim2006}. The majority of the heat generated in this process is carried away with the chip \citep{Shaw2005}.  In order to form the chip, the cutting tool must have a hardness higher than the hardness of the workpiece material. In addition, the cutting tool material must possess a combination of: (i) high-temperature hardness and chemical stability (ii) abrasive wear resistance (iii) resistance to brittle fracture. Tool wear is defined as a gradual loss of the tool material in the tool-workpiece contact zones \citep{Klocke1999}. When the quality of the tool cutting edge is lost because of wear or breakage, the tool must be replaced.  Cutting tools experience various modes of wear such as: flank, notch, crater wear and edge chipping \citep{Altintas2012}. These modes result from a combination of wear mechanisms, some of which include: adhesion, abrasion, oxidation, diffusion and fracture wear. Figure~\ref{fig:Characteristic tool wear patterns} shows some examples of this type of wear.
 
	The tool wear rates depend on a range of fixed process  and control parameters. Taylor's tool life equation is an example of a simple model describing the relationship between tool life and the cutting conditions \citep{Altintas2012}:
$$v_c T^n=C$$
 where $v_c$ is the cutting speed, $T$ is the tool life, $n$ and $C$ are experimentally determined material constants \citep{Shaw2005}. The cutting speed and the chip thickness are thus indicated to be the main cutting conditions that affect the tool wear rates. A significant amount of research has focused on optimisation of control parameters given combinations of the workpiece material, machine and cutting tool-characteristics. The process constraints and costs associated with the tool-wear are further aggravated when machining difficult to cut alloys \citep{Shaw2005}. Machinability of an alloy is subject to its characteristics, where titanium, nickel and cobalt based alloys are widely regarded as DTC. In previous work \citep{Baron2015c} we compared Co-Cr-Mo alloy with Ti-6Al-4V and subsequently classified it as DTC. Cobalt and nickel based alloys contain highly abrasive, carbide particles generating abrasive wear \citep{Shaw2005}. The relatively low thermal conductivity of these alloys and high work hardening rates results in heat generation, resulting in temperatures of up to 1200$^\circ$C and high magnitudes of stresses at the tool-workpiece interface \citep{Altintas2012, Settineri2008}. These conditions promote oxidation and diffusion of tool material. Common countermeasures applied to carbide tooling are PVC or CVD coatings of TiN or TiAlN. More recent efforts also include (i) rake and/or flank face structuring (ii) cryogenic machining \citep{Altintas2012} (iii) nanostructure and superlattice PVD single and multi-coating layer coatings \citep{Uhlmann2009}.

	The advances outlined above focus on either prolonging the tool-life or increasing the material removal rates but do not advance in the `levels of intelligence' (Figure~\ref{fig:Roadmap}). As the tool-life is often application specific, the majority of production chains continue to utilise machining-time tool change criteria \citep{Altintas2012}. Measurement of the cutting force is an example of indirect, in-process measurement while periodic tool edge inspection is an example of direct, intermittent tool condition monitoring (TCM) strategy \citep{Byrne20161}. Incorporation of a machining process TCM system for an unattended manufacturing process relies on sensors that measure a range of in-process variables influenced by the condition of the cutting tool. The output of the sensors is subject to analogue and/or digital signal conditioning followed by signal processing in order to generate parameters that are correlated with the state of the cutting tool \citep{Teti2010}. These parameters are an input to a decision making support system for final interpretation. The output from the decision making system can be communicated to the machine tool operator or fed to the CNC controller for execution \citep{Teti2010}. 

	Models to predict the tool-wear in the time-domain include auto regressive, moving average and auto-regressive moving average \citep{Teti2010, Dimla2000, Abouelatta2001}. Advancement in `levels of intelligence' and incorporation of tool TCM system in unattended manufacturing process requires effective decision making support structure. The cognitive systems most commonly employed for this purpose include neural networks, fuzzy logic, generic algorithms and hybrid systems \citep{Byrne1995, Teti2010}.

	In this paper, we employ a Bayesian approach to statistical modelling of tool wear. The Bayesian approach admits the incorporation of expert prior knowledge into the estimation of model parameters. This is implemented by the specification of prior distributions. A comprehensive explanation of Bayesian data analysis is provided in \cite{Gelman2003}. As part of our modelling approach, we use non-parametric Bayesian models known as Gaussian processes. These allow us to specify an infinite dimensional Gaussian distribution which can flexibly adapt to observed data. This technique is utilised here to link model parameters under different experimental conditions by borrowing strength between experiments. Posterior predictive distributions of the Gaussian Process are used to estimate the model parameters for experimental conditions not used in the study.

	In Bayesian analysis, prior distributions are combined with the experimental data to produce a posterior distribution of the parameters given the data. It is often difficult to perform inference based on this distribution directly since the distribution itself is often intractable. Instead, inference is done via a Markov chain Monte Carlo (MCMC) algorithm that draws samples from the posterior distribution. More information on MCMC algorithms can be found in \cite{Brooks2011}. Much development has taken place in MCMC algorithms over the last 20 years. One recent major development was Hamiltonian Monte Carlo (HMC) algorithms \citep{Neal2011, Girolami2011, Hoffmann2014}. These algorithms make intelligent proposals based on first order gradient information. We use the HMC algorithm implemented in the Stan software \citep{Carpenter2016} which is interfaced with R using the {\tt rstan} package. 

	\section{Methods}
    \label{sec:methods}
    
    \subsection{Experimental Plan and Design}

\begin{figure}[h]
      \centering
      \includegraphics[width=0.9\textwidth]{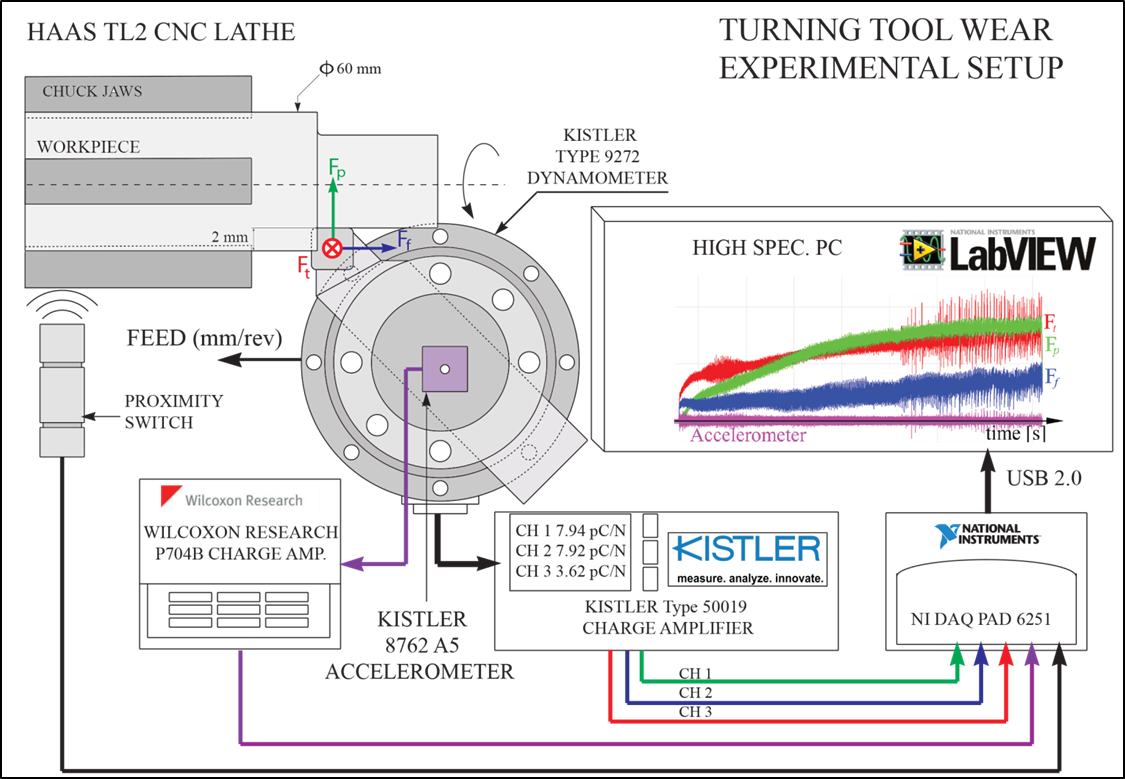}
      \caption{Configuration of the tool-wear experimental setup.}
      \label{fig:ExpConfig}
    \end{figure}

\indent We perform single point turning tool-wear experiments to examine the influence of the process control parameters. We vary cutting speed $v_c$ (m/min) and feed-rate $f$ (mm/rev), and evaluate their effects on the tool-wear rates and resultant forces in the tangential force, $F_t$, feed force, $F_f$, and axial (passive) force, $F_p$, directions (Figure~\ref{fig:ExpConfig}). It should be noted here that the feed rate $(\mu m/rev)$ in turning is comparable to the more basic undeformed chip thickness referred to in analysis of cutting processes. The experimental parameters were based on on our previous research into orthogonal cutting of Co-Cr-Mo alloy ASTM F1537 standard reported in \cite{Baron2015b}. 

\begin{figure}[h]
	\centering
	\includegraphics[width=0.8\textwidth]{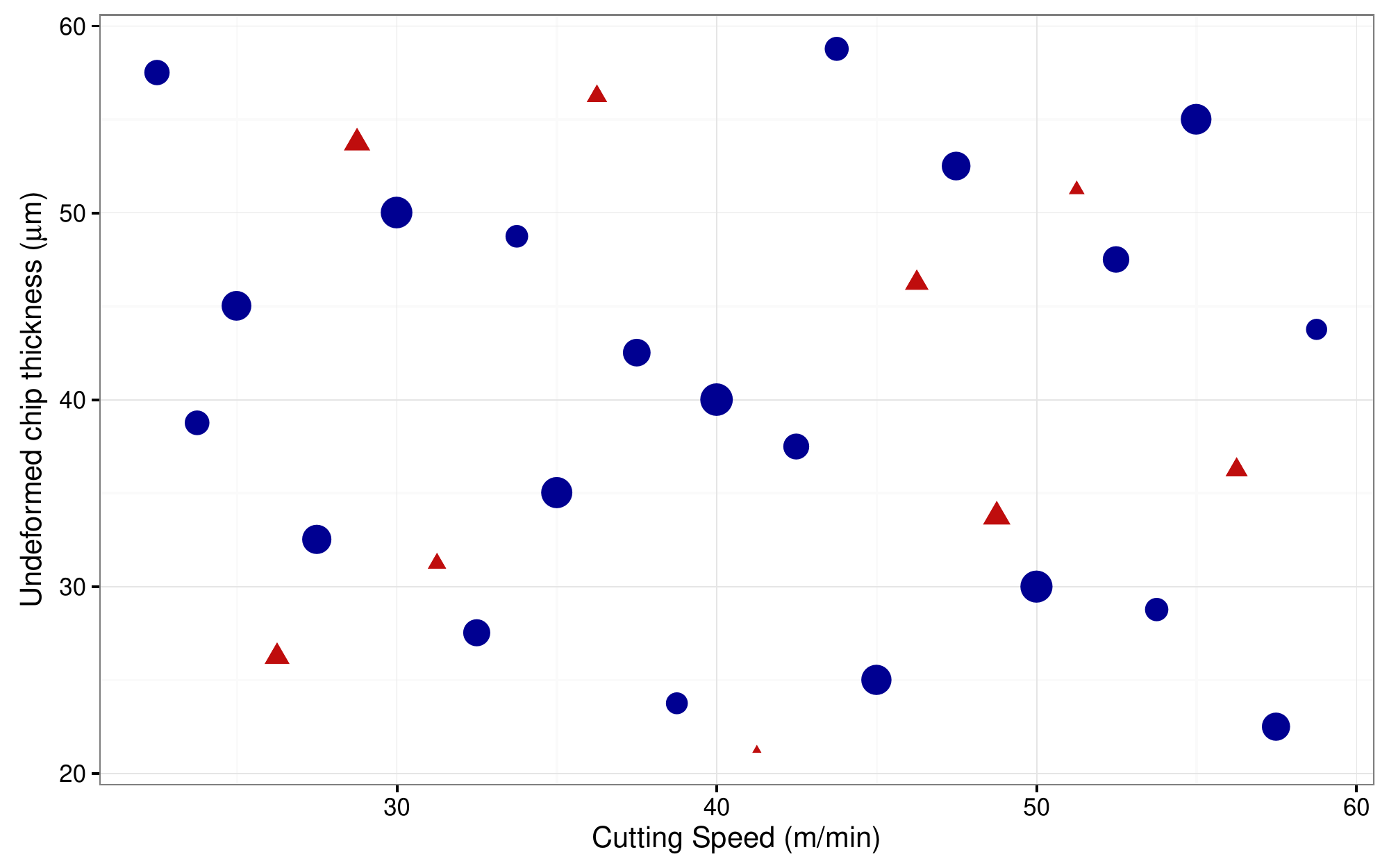}
	\caption{Experimental design layout following a Sobol sequence. The larger points denote experimental settings considered first while settings tested later are smaller. The 21 blue discs denote settings that were actually used while the 9 red triangles denote the following points in the sequence for which there was not time to conduct the experiments. It can be seen that the sequence provides initially coarse coverage that gets progressively finer.}
	\label{fig:Experiment_design}
      \end{figure}

	A full factorial randomised design over the feasible ranges for cutting speed and feed-rate was set-up using a Sobol sequence \citep{Sobol1967}. This sequence seeks to minimise any gaps or holes in a multidimensional space whilst simultaneously minimising any holes in projections of the sequence into a lower number of dimensions. The sequence initially contains points that cover the range of settings in a coarse manner and, as the sequence progresses, points are added to obtain a finer degree of coverage. This behaviour supports progressive augmentation of test settings \citep{Azmin2015} thus allowing for the maximum number of tests in the time available to complete the experiment. Figure~\ref{fig:Experiment_design} shows the first 21 prioritised test settings that were used in the experiment plus the following 9 test settings that could have been used to augment the data if time had allowed. The tests are run at the settings corresponding to the larger points first, moving to the smaller points afterwards. It is clear that the sequence provides an initially coarse coverage that becomes progressively finer. This design enables estimation of non-linear effects and interactions between control variables, whilst maximising the time value of the experiment.

    \subsection{Experimental Procedures}
    
   The tests were carried out on the HAAS TL2 CNC lathe with an 8.9 kW spindle motor rated for maximum speed of 2000 RPM and driving a 208mm capacity chuck. The forces, $F_t$, $F_p$, and $F_f$ were recorded using a force measuring system comprised of Kistler Type 9272 dynamometer, a Kistler Type 5019 charge amplifier. The dynamometer was secured on the machine tool using a rigid fixture system. The fixture-dynamometer setup was analysed using FEA for stiffness in the horizontal and the vertical planes (found to be 148 and 181 kNmm$^{-1}$ respectively). A modal analysis of this assembly, assuming infinite tool-base rigidity showed lowest modal frequency to be 3.6 kHz, well above the in-plane dynamometer natural frequency of 1.5kHz.  Prior to the experiment, the force measuring system was cross calibrated off the machine on a Hounsfield tensometer. It was found to conform within 5\% of the Hounsfield measurement over the range from 0 to 4 kN. Tool vibration was monitored using a Kistler 8762A5 accelerometer with Wilcoxon Research P704B charge amplifier. A proximity sensor was installed on the machine to allow for accurate determination of the number of rotations and the rotational speed of the chuck. It therefore allowed for verification of the process control parameters. The output of the proximity switch, force, and vibration measuring system was recorded using NI DAQ PAD 6051 and LabView software with the sampling rate set to 150 kHz. The schematic of the experimental setup is shown in Figure~\ref{fig:ExpConfig}. The force data was post-processed and analysed using NI DiaDem2014 software which included a low-pass digital filtering of 1.5 kHz in order to remove any influence of dynamometer resonant frequency of 1.5 kHz \citep{Kistler}. Modal analysis was used to determine the work offset length to ensure stability. As a result a round Co-Cr-Mo ASTM F75 bar of 61mm diameter was secured in the chuck such that the work offset, measured from the chuck, was 120mm. The chemical composition of the workpiece is given in Table~\ref{tab:ChemComp}.
   
    \begin{table}[h]
      \centering
      \caption{Chemical composition requirements of alloy conforming to ASTM F75 standard.}
      \begin{tabular}{lcccccccc}
	\hline\noalign{\smallskip}
	Material & Cr	 & Mo & Ni & Fe & C &	Si &	Mn &	 Co\\
	\noalign{\smallskip}\hline\noalign{\smallskip}
	ASTM F75 \citep{ASTM} &  27-30 & 5-7 & $<$0.5 & $\leq$0.75 & $<$0.14 & $<$1 & $<$1 & Bal.\\
	\noalign{\smallskip}\hline
      \end{tabular}
      \label{tab:ChemComp}
    \end{table}

The workpiece material nominal properties, listed in Table~\ref{tab:TestParams}, were validated in part by hardness measurements taken at intervals in the radial and axial directions where the mean measured hardness is given in Table \ref{tab:MechProps}. 

    \begin{table}[h]
      \centering
      \caption{Material nominal properties of an alloy conforming to ASTM F75 standard.}
      \begin{tabular}{lcc}
	\hline\noalign{\smallskip}
	& Units & ASTM F75 \citep{Kaiser2013}\\
	\noalign{\smallskip}\hline\noalign{\smallskip}
	Tensile Strength & MPa &	920\\
	0.2\% Proof Stress & MPa &	527\\
	Elongation & \% &	16.5\\
	Young's Modulus & GPa & 216\\
	Hardness & HRC & 34\\
	Density ($\rho$) & Kg.m$^{-3}$	 & 8,300\\
	Thermal Diffusivity & m$^2$.s$^{-1}$ & $3 \times 10^{-6}$\\
	\noalign{\smallskip}\hline
      \end{tabular}
      \label{tab:MechProps}
    \end{table}
    
    \begin{table}[h]
      \centering
      \caption{Detailed list of fixed experimental parameters.}
      \begin{tabular}{lll}
	\hline\noalign{\smallskip}
    Test Parameter & Units: & Value:\\
	\noalign{\smallskip}\hline\noalign{\smallskip}
	Work Material &&\\ 
	\noalign{\smallskip}\hline\noalign{\smallskip}
	Material properties: & &		See Table~\ref{tab:MechProps}\\
	Stock Shape Dimensions & mm & ASTM F75 $\emptyset$ 62  Round bar\\
	Measured Hardness & HV(HRC) & 322(34) for ASTM F75\\
	Depth of cut	& mm & 1.9\\
	\noalign{\smallskip}\hline\noalign{\smallskip}
	Tool Insert	 &&\\
	\noalign{\smallskip}\hline\noalign{\smallskip}
	Insert Code: & N/A & SCMW 120408 H13A\\
	Rake Angle: & $^{\circ}$ &	0\\
	Relief Angle: & 	$^{\circ}$ & 7\\
	Nose Radius & $mm$ & 0.8 \\	
	Cutting Edge Radius, $r_\beta$: & $\mu$m & 14 \\
	\noalign{\smallskip}\hline\noalign{\smallskip}
	Cutting Fluid	 &&\\
	\noalign{\smallskip}\hline\noalign{\smallskip}
	Cutting Fluid Type:	& N/A & FUSCH ECOCOOL ULTRALIFE A\\
	Cutting Fluid Flow:	& l.min$^{-1}$ & 2.5\\
	Concentration: & \% & 8.50\\
	\noalign{\smallskip}\hline
      \end{tabular}
      \label{tab:TestParams}
    \end{table}
    
	The test procedure involved running an outside-diameter turning operation with the depth of cut equal to 1.9 $mm$. The tool moved in the axial direction at a set feed-rate $(\mu m/rev)$ with the chuck rotating at fixed rpm such that the pre-determined cutting speed was achieved. The wear scar on the rake and the flank face of the insert was first photographed and measured using a Keyence VHX-2000 optical microscope at intervals corresponding to $\approx$ 20 $m$ length of the spiral of removed material. The criterion for determining the tool life (m), the value of cutting length when the fool failure occurs, was either a visible increase in the variance of $F_t$ or a corresponding flank wear value greater than 200$\mu m$ in cases where a clear change in  the variance of $F_t$ was not visible. In general, there was a visible change in $F_t$ for higher cutting speeds while for lower cutting speeds the tool-wear was more gradual. It must be noted that the flank wear criterion could be exceeded before the onset of the force variance criterion. In the case of gradual tool wear, the length of cut corresponding to the closest value of 200$\mu m$ flank wear was selected as the value of tool life. 

    \subsection{Statistical Methods}
      \label{subsec:statmethods}
      
	We use a Bayesian Hierarchical Model \citep{Gelman2006b} to simultaneously learn the tool wear rates in each force direction whilst constraining them in an $\mathbb{R}^2$ random field according to their cutting speeds and feed rate. The Bayesian approach allows us to borrow strength between experiments by using informative prior distributions on the latent parameters of interest, and thus to improve the precision of our estimates. The top level of our model (the likelihood) is a simple linear regression of force against length of cut for each experiment. At the second level the latent slopes (representing e.g. the feed force) are given a non-parametric prior which regularises their values and allows us to predict feed force for experiments not yet run. 

	We start by defining the pair $(F_{ij}, L_{ij})$ to be respectively the force (either $F_t$, $F_f$ or $F_p$) and length measurements for experiment $i$ ($i=1,\ldots,K$ experiments) and measurement $j$ ($j=1,\ldots,N_i$ measurements for experiment $i$). The top level of the model is then:
      $$F_{ij} = \alpha_i + \beta_i L_{ij} + \varepsilon_{ij}$$
where $\alpha_i$ and $\beta_i$ are the intercept and slope parameters respectively for experiment $i$, and $\varepsilon_{ij} \sim N(0,\sigma_i^2)$ is a residual term.

	At the second level of the model the slopes $\beta_i$ are given a Gaussian Process (GP) prior \citep{Rasmussen2006}, such that:
$$\beta \sim GP(\mu(.), \Sigma(.,.))$$
where $\mu(.)$ indicates the mean function, here set to constant $\mu_\beta$, and $\Sigma(.,.)$ represents the autocovariance function. We use an isotropic Gaussian autocovariance such that:
$$\Sigma_{mn} = \eta_\beta^2\exp\left[-\rho_1(V_{m} - V_{n})^2 - \rho_2(f_m - f_n)^2\right]$$
where $V_m$ and $f_m$ are the cutting speed and feed rate for experiment $m$. The additional parameters $\eta_\beta^2$, $\rho_1$ and $\rho_2$ control the degree of smoothing across the $\mathbb{R}^2$ field; all are restricted to be positive to ensure positive definiteness of the resulting covariance matrix. A similar restriction requires $\Sigma_{mm} = \eta_\beta^2 + \sigma_\beta^2$ where $\sigma_\beta^2$ represents experimental replicate variance. When the values $\rho_1$ and $\rho_2$ are small then the autocovariance decreases slowly with distance and the field is smooth, and vice versa. The $K \times K$ matrix $\Sigma$ forms a valid covariance matrix for the multivariate Gaussian distribution that is used to fit the model. 

	The model nuisance parameters $\alpha_i$ are also at the second level. They are regularised such that $\alpha_i \sim N(\mu_\alpha, \sigma_\alpha^2)$. This enables further borrowing of strength between experiments. At the lowest level the hyper-parameters $\Theta = \{ \sigma, \sigma_\alpha, \mu_\alpha, \mu_\beta, \eta_\beta, \sigma_\beta, \rho_1, \rho_2\}$ are constrained using weakly informative prior distributions. Informative prior distributions are to be preferred (such as the GP prior in the second level), but hyper-parameters for GP models are notoriously difficult to obtain so we follow the weakly-informative convention of \cite{Gelman2006a}. The specific priors used for the experiment described in this paper are detailed in Section~\ref{sec:results}. 

Taken together we form a posterior distribution:
\begin{eqnarray}
p(\Theta,\beta|L, F) &\propto& p(F|L, \alpha, \beta, \sigma) p(\beta| \mu_\beta, \eta_\beta, \sigma_\beta, \rho_1, \rho_2) p(\alpha|\mu_\alpha,\sigma_\alpha) p(\Theta)
\label{post}
\end{eqnarray}
After fitting we can estimate the predictive distribution of rates $\beta^*$ at new values $V^*$ and $f^*$ via the GP identity:
\begin{eqnarray}
\beta^*|\beta, \Theta &\sim& N(\beta + \Sigma_\beta^* \Sigma_\beta^{-1} (\beta - \mu_\beta), \eta_\beta^2 + \sigma_\beta^2 - \Sigma_\beta^* \Sigma_\beta^{-1} (\Sigma_\beta^*)^T)
\label{predictive}
\end{eqnarray}
where $[ \Sigma_\beta^* ]_{m} = \eta_\beta^2\exp\left[-\rho_1(V^* - V_{m})^2 - \rho_2(f^* - f_m)^2\right]$. For a proof see \cite{Krzanowski2000}. We take account of parametric uncertainty by integrating over the other parameters:
$$p(\beta^*|F, L) = \int p(\beta^*|\Theta,\beta) p(\Theta,\beta|L, F) \partial \Theta \;\partial \beta$$
where the first term in the integrand is the GP identity (Equation \ref{predictive}) and is the second is the posterior distribution  (Equation \ref{post}).

	We fit the model using the probabilistic programming language Stan \citep{Carpenter2016} through the R package {\tt rstan}. The package uses Hamiltonian Monte Carlo, a variant on the standard Markov chain Monte Carlo \citep{Brooks2011}, which samples from the posterior distribution of all parameters simultaneously using the derivatives of the posterior up to proportionality. The particular method used by {\tt rstan} is called the No U-turn Sampler \citep[NUTS;][]{Hoffmann2014} which avoids over-shooting when sampling parameters in particular directions. In contrast to other HMCMC algorithms, the NUTS algorithm does not require the user to specify the step size or the number of steps and thus requires no hand-tuning.

	\section{Results}
    \label{sec:results}

    Twenty one tests were carried out under various settings for cutting speed and feed rate. The exact control parameters used for the experiments are detailed in the Supplementary Material. In general tool life ranged from 10 $m$ to 255$m$, with Test 7 ($v_{c}=20$m/min, $f=45 \mu$m/rev) showing the longest tool life (255$m$) and, correspondingly, Test 10 ($v_{c}=58$ $m/min$, $f=22.5\mu m$/rev) giving the shortest (10$m$). As would be expected, higher speeds and larger values for feed rate generally lead to shorter tool life.
   Given workpiece dimensions, most tests required multiple cutting passes before the tool was worn out. Thus the data obtained from the dynamometer included periods where the toll was not in contact with the workpiece. Changepoint analysis using binary segmentation \citep{Killick2014, Scott1974} was used as a guide to determine the contact phases. These phases were then concatenated before the model described above was fitted.
    The regression model with Gaussian process prior for the slope parameters is fitted for each of the three forces. In order to estimate the model in a Bayesian framework, prior distributions must be specified. The following vague prior distributions were used in the analysis of these data:

    \medskip
    \begin{tabular}{cc}
      \begin{minipage}{0.5\textwidth}
	\begin{itemize}
	  \item $ \sigma_i^2 \sim Half-Cauchy(0, 10)$
	  \item $ \mu_\alpha \sim N(0, 10^2)$
	  \item $ \sigma^2_\alpha \sim Half-Cauchy(0, 10)$
	  \item $\mu_\beta \sim N(0, 10^2)$
	\end{itemize}
      \end{minipage}
      &
      \begin{minipage}{0.5\textwidth}
	\begin{itemize}
	  \item $ \sigma_\beta^2 \sim Half-Cauchy(0, 5)$
	  \item $ \eta_\beta^2 \sim Half-Cauchy(0, 5)$
	  \item $ \rho_1^{-1} \sim Half-Cauchy(0, 5)$
	  \item $ \rho_2^{-1} \sim Half-Cauchy(0, 5)$
	\end{itemize}
      \end{minipage}
    \end{tabular}
    \medskip
    
The convergence of the Markov chains was assessed by examining trace plots of the sampled values and the Potential Scale Reduction Factor \citep[PSRF;][]{Gelman1992, Brooks1998}. After running the model all values of PSRF were close to 1, which is the desired value for satisfactory convergence.

    The parameter estimates were obtained from the posterior probability distribution of the HMCMC samples from the given the data as in Equation \ref{post}. The estimated rates of change of force over cutting distance are reported in the Supplementary material for each of the three forces, $F_f$, $F_t$ and $F_p$, for each of the 21 experiments.

    The obtained posterior parameter values can be used to estimate the rate of change of feed force for any combination of feed rate and cutting speed (within a neighbourhood of the observed values) via Equation \ref{predictive} on a regular grid. Slope estimates corresponding to unobserved experimental settings are naturally formed via this multivariate Gaussian conditional distribution given the twenty one latent slope estimates. Standard deviations of these estimates can be obtained similarly. Further information on Gaussian processes for fitting and prediction can be found in \cite{Rasmussen2006}.
    
\begin{figure}[!p]

    \begin{subfigure}[b]{0.49\textwidth}
      \includegraphics[width=0.9\textwidth]{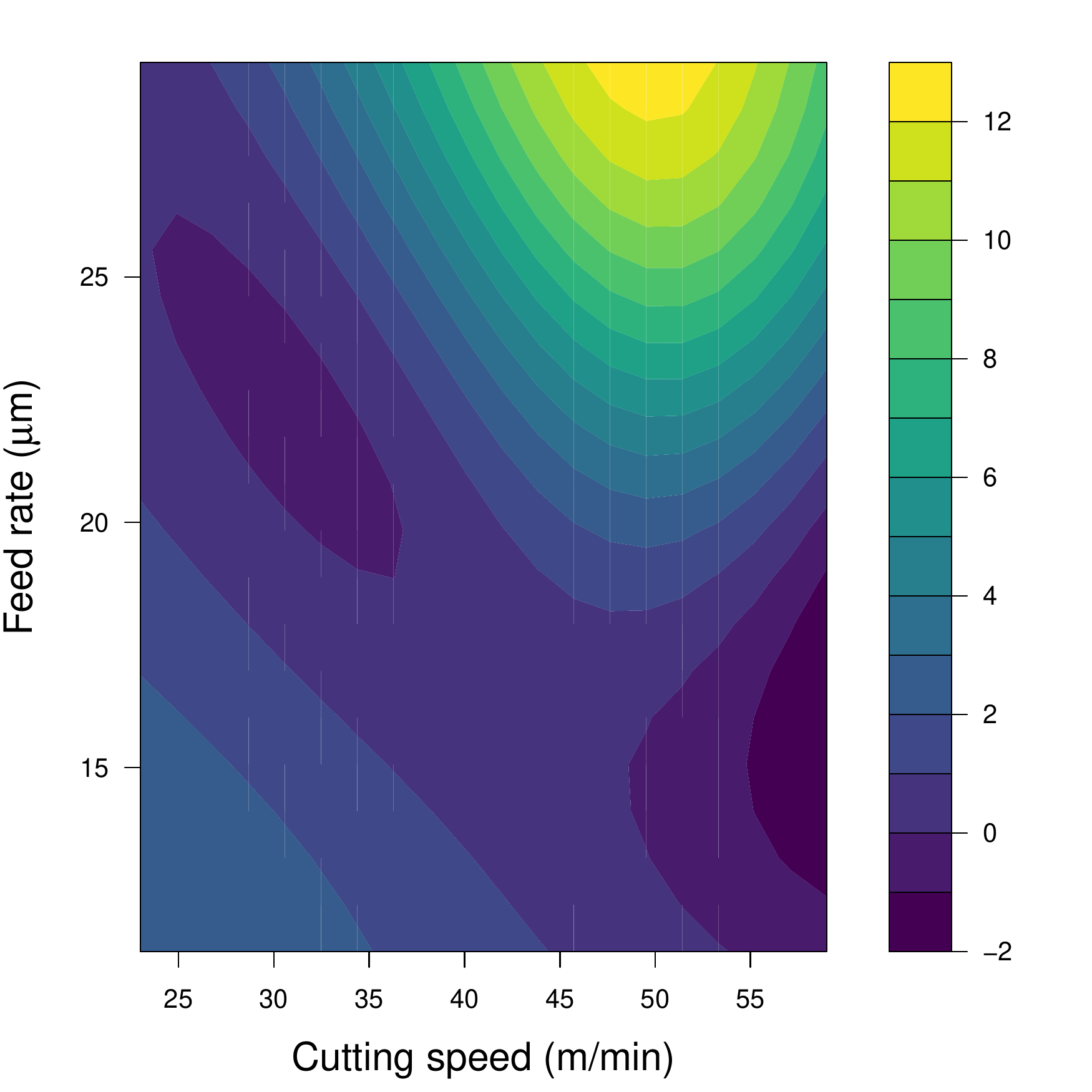}
      \label{fig:Fy_contour}
    \end{subfigure}
    ~
    \begin{subfigure}[b]{0.49\textwidth}
      \includegraphics[width=0.9\textwidth]{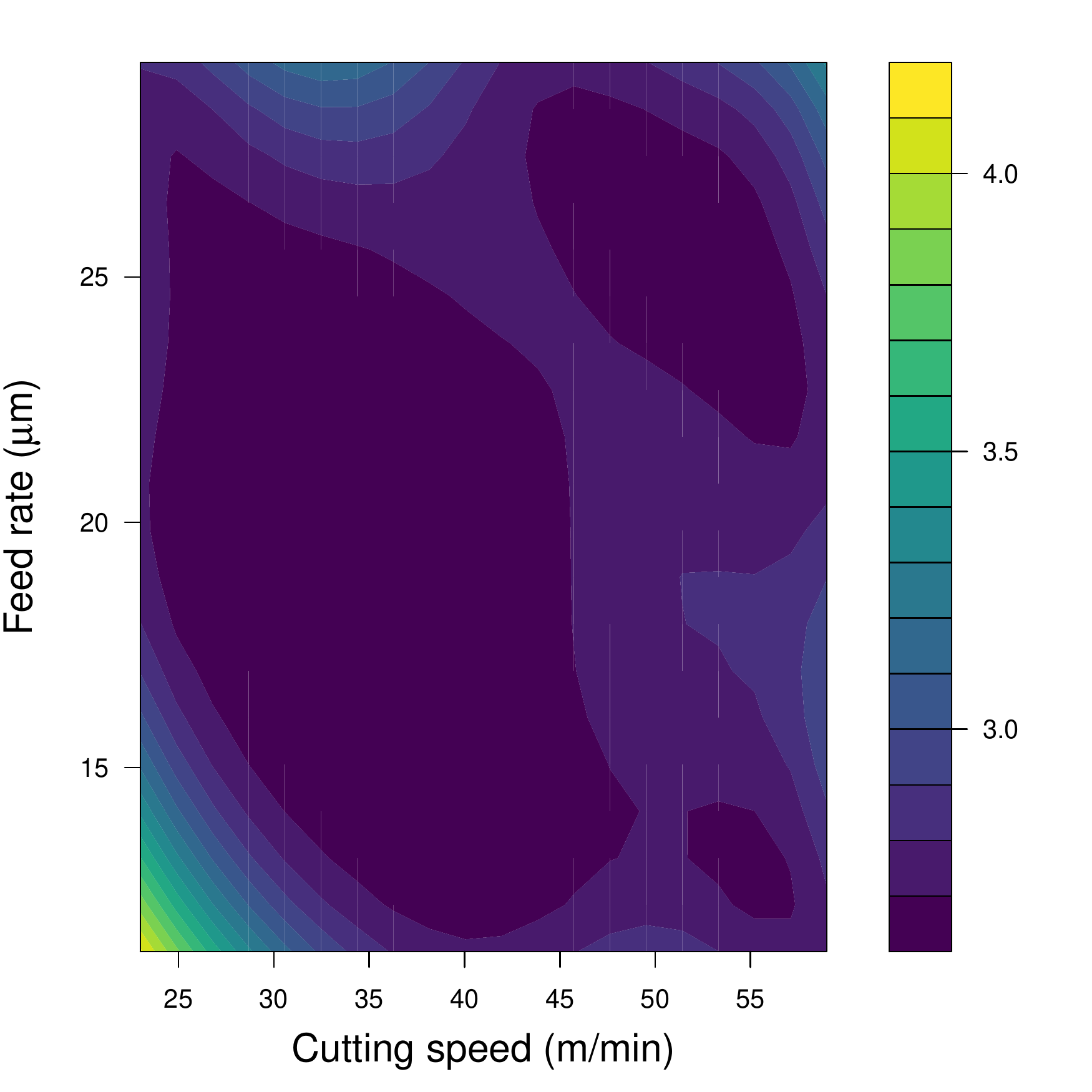}
      \label{fig:Fy_sd}
    \end{subfigure}
    \\
      \smallskip
    \begin{subfigure}[b]{0.49\textwidth}
      \includegraphics[width=0.9\textwidth]{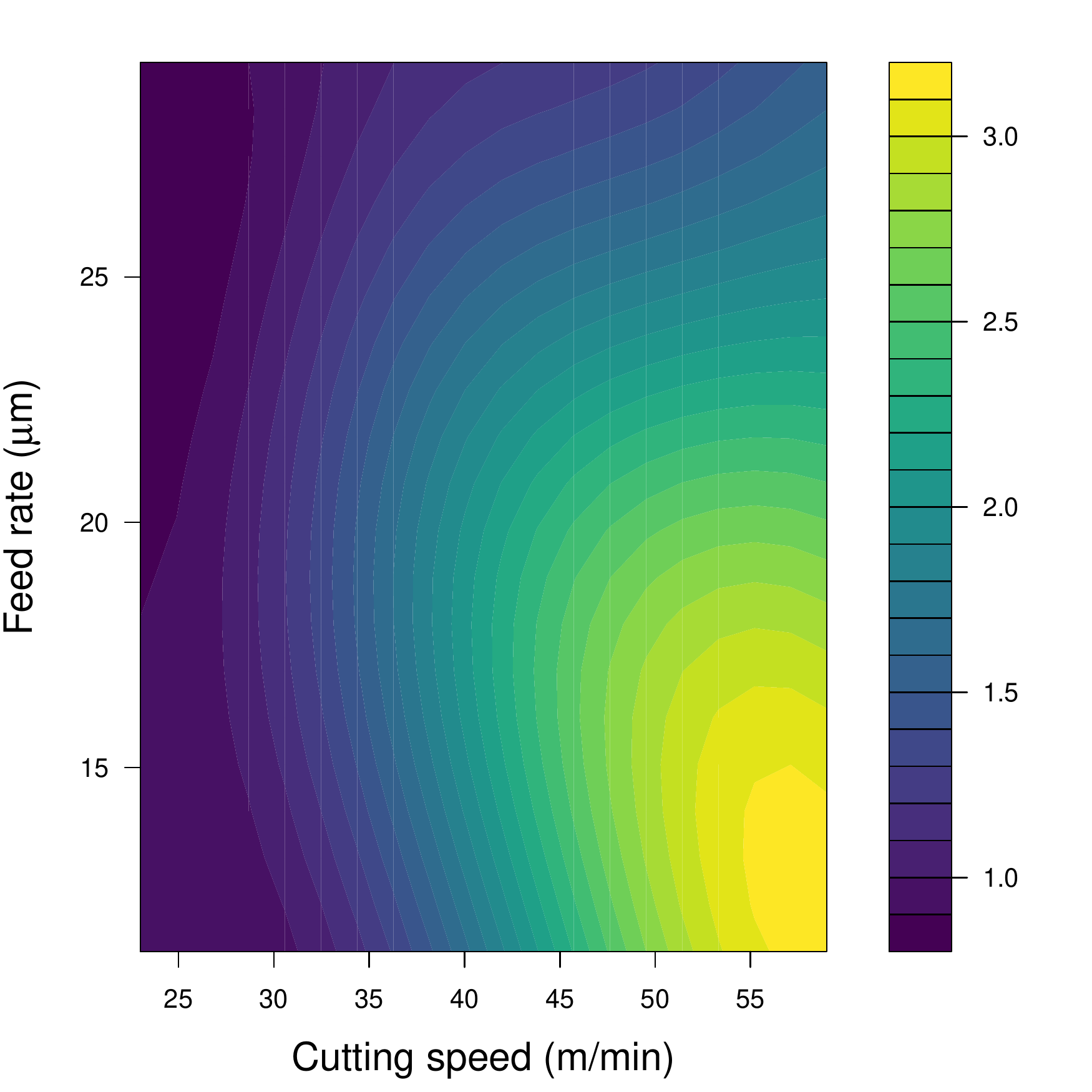}
      \label{fig:Fx_contour}
    \end{subfigure}
    ~
    \begin{subfigure}[b]{0.49\textwidth}
      \includegraphics[width=0.9\textwidth]{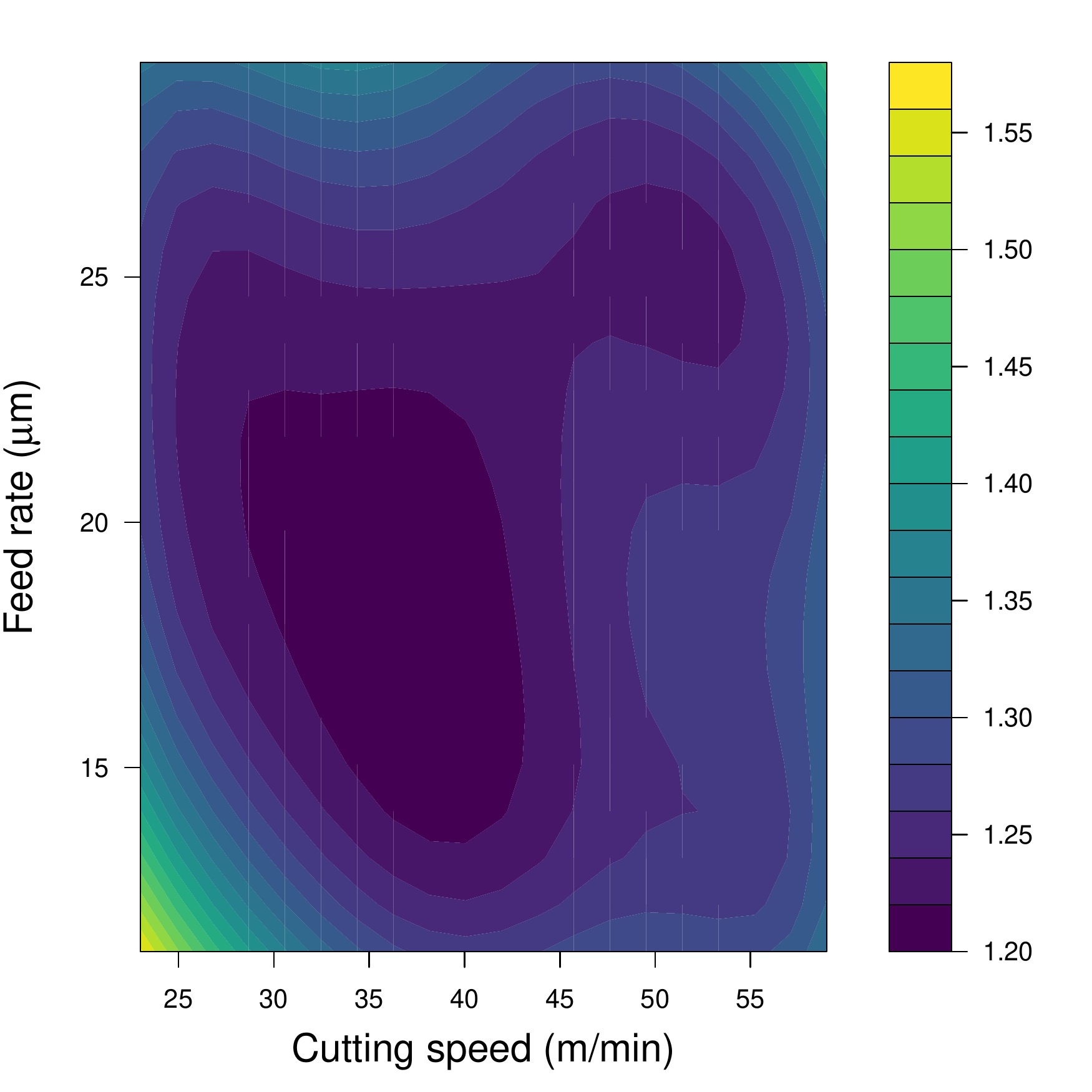}
      \label{fig:Fx_sd}
    \end{subfigure}
    \\ 
     \smallskip
    \begin{subfigure}[b]{0.49\textwidth}
      \includegraphics[width=0.9\textwidth]{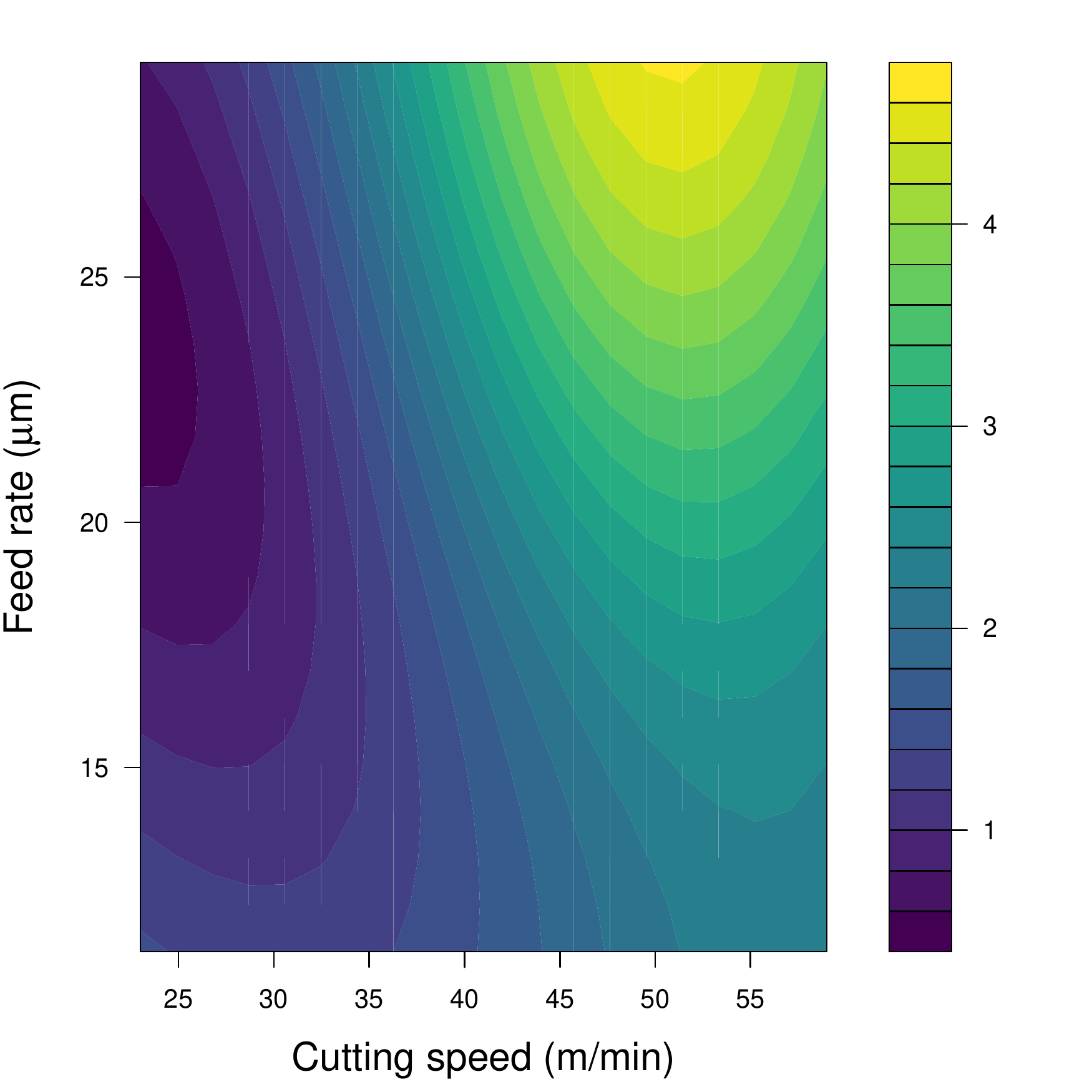}
      \label{fig:Fz_contour}
    \end{subfigure}
    ~
    \begin{subfigure}[b]{0.49\textwidth}
      \includegraphics[width=0.9\textwidth]{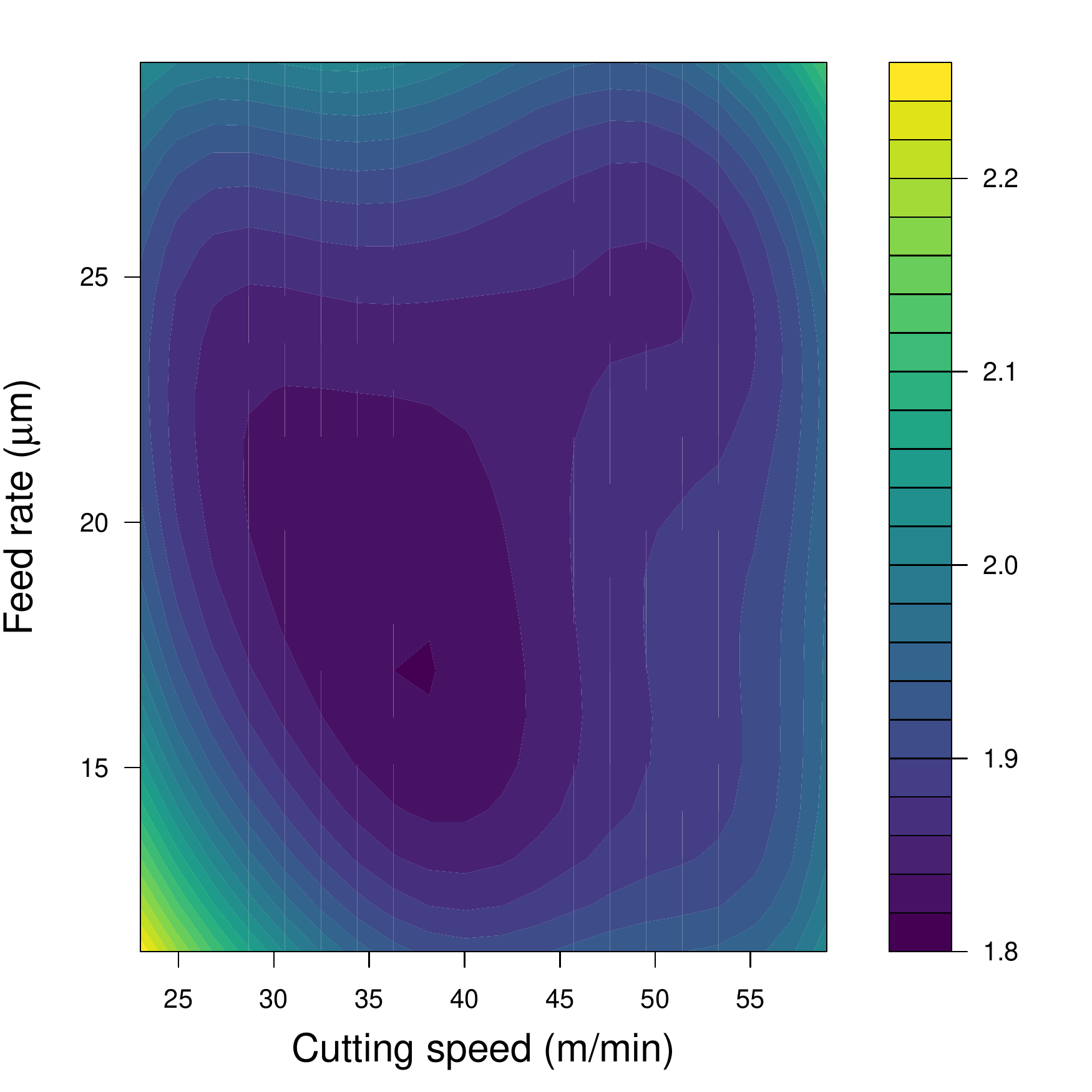}
      \label{fig:Fz_sd}
    \end{subfigure}
    \caption{The first column shows the estimated mean rates of changes whilst the second shows the estimated standard deviations. Note the differing scales. The rows correspond to the rate of change of feed force ($F_f$), passive force ($F_p$) and tangential force ($F_t$) respectively. Each surface is a function of feed rate and cutting speed.}
    \label{fig:surfaces}
\end{figure}

    The first columns of Figure~\ref{fig:surfaces} shows surface plots for the estimated rates of change in each of the three forces varying by speed and thickness. A grid of 400 points was formed on the illustrated region and the rate of change in  force was estimated at each point using the Gaussian process. Standard deviations of these estimates are shown in the surface plots given in the second column.
    
\begin{figure}[!h]
    \centering
    \begin{subfigure}[b]{0.49\textwidth}
      \includegraphics[width=0.9\textwidth]{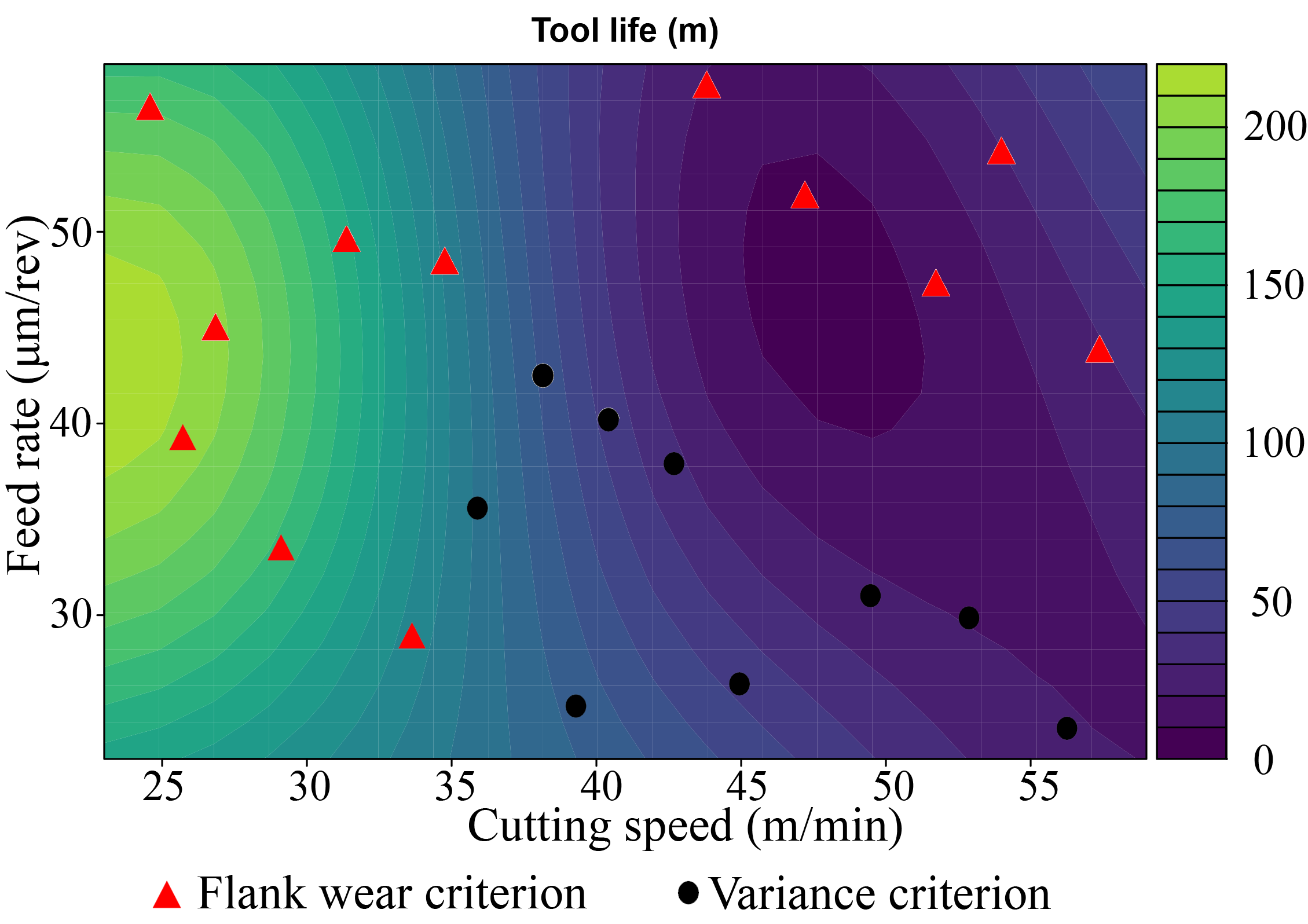}
\caption{}
      \label{fig:LifeSurface}
    \end{subfigure}
    ~
    \begin{subfigure}[b]{0.49\textwidth}
      \includegraphics[width=0.9\textwidth]{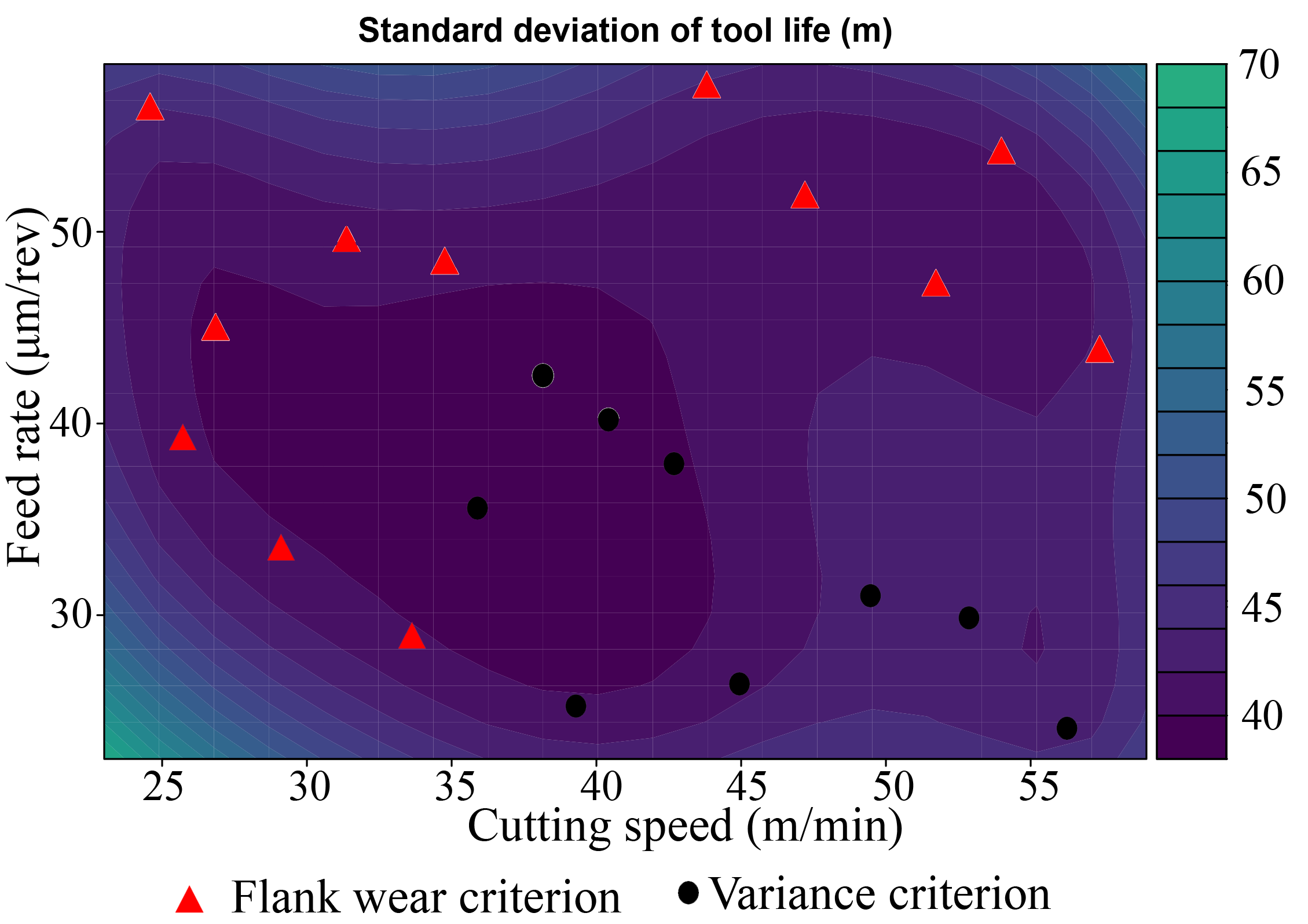}
\caption{}
      \label{fig:LifeSurface_sd}
    \end{subfigure}
    \caption{The first column shows surface plots indicating the mean tool life (in metres) as a function of feed rate and cutting speed. The second column shows surface plots indicating the standard deviation of these estimates. The tool end-of life criteria are superimposed on both surface plots.}\label{fig:LifeSurfaces}
\end{figure}

	In a separate analysis, a similar Gaussian process model was used to relate tool life (m) to the two variables of interest, cutting speed and feed rate. This model did not require the top level linear regression step. Once again, the NUTS algorithm was used to fit the model. Figure~\ref{fig:LifeSurface} shows a surface plot illustrating tool life as a function of $v_c$ and $f$. Figure~\ref{fig:LifeSurface_sd} shows the standard deviations of these estimates.

The response surface plots (Figure~\ref{fig:surfaces}) map out the rates of change of the force components as a function of the most fundamental cutting parameters. The top left and bottom left panels demonstrate a clear correlation between the rates of change of the tangential ($F_t$) and feed force ($F_f$) components, while the mapping of the passive force $F_p$ component rate of change is characteristically different. This is attributed physically to the kinematics and the tool geometry in turning where the magnitude of $F_t$ and $F_f$ are both determined by the removal of most material on the side edge (see Figure~\ref{fig:ExpConfig}) of the tool in the feed direction. The corollary is that the nose radius is primarily subjected to lower levels of locally varying undeformed chip thickness with mechanisms of surface plastic deformation and elastic friction applying without material removal \citep{chou2004tool}. Thus the rates of change in $F_t$ and $F_f$ are expected to be related at least to the measured flank wear levels. This is indeed surmised by the inverse correlation of the tool life as presented in Figure~\ref{fig:LifeSurfaces}. Nonetheless the rates of change in $F_p$ are expected to be indicative of other (potentially adverse) effects in turning, for example changes in surface finish and integrity \citep{pawade2007investigation}. 

We conclude that the rates of change in $F_t$ and $F_f$ correlate with tool-wear rate and provide useful surrogate measures. The response surface plots shown enable the determination of parameter ranges that result in maximum tool life and also prediction by interpolation of results for values not specifically set-up in the experiment (but within the range examined). For example, the use of low cutting speeds and high feed rate will extend tool life while maintaining high material removal rates (higher feed rate decreases the surface roughness but this is not considered here). The results also provide a baseline for comparison of future research on tool-wear in turning and milling of ASTM F75 Co-Cr-Mo and can be used for optimisation of the manufacturing process. This methodology can be further utilised to compare the tool life for different cutting inserts. It should be noted however, that this model should not be used to estimate the rate of change in cutting forces by extrapolating far outside the range of cutting speeds and feed rates considered in this experiment. A further experiment, at low $v_c$ and $f$ settings, would provide a more detailed description of the profile of the components of the cutting force and progression of the tool wear.  

  \section{Discussion}
    \label{sec:discussion}
   
We have presented the results of a novel statistical approach to experimental design applied to machining processes of tool wear progression in turning of medical grade Cobalt Chromium Molybdenum Alloy (ASTM F75). The Sobol sequence design methodology employed allows for the maximum number of tests to be carried out in a designated period of time while also covering the entire parameter range under investigation. We believe it has potential benefits to industry, particularly in time consuming tool-wear trials. While tool wear studies have employed data analytic techniques, to the best of our knowledge the use of Gaussian processes in the manner presented here is novel in this area. The change-point detection technique used allows for filtering out of non-contact and ramp-up stages of the force measurements in this type of experiment and has potential for implementation in indirect, in-process TCM systems.

\begin{figure}[ht]
      \centering
      \includegraphics[width=0.75\textwidth]{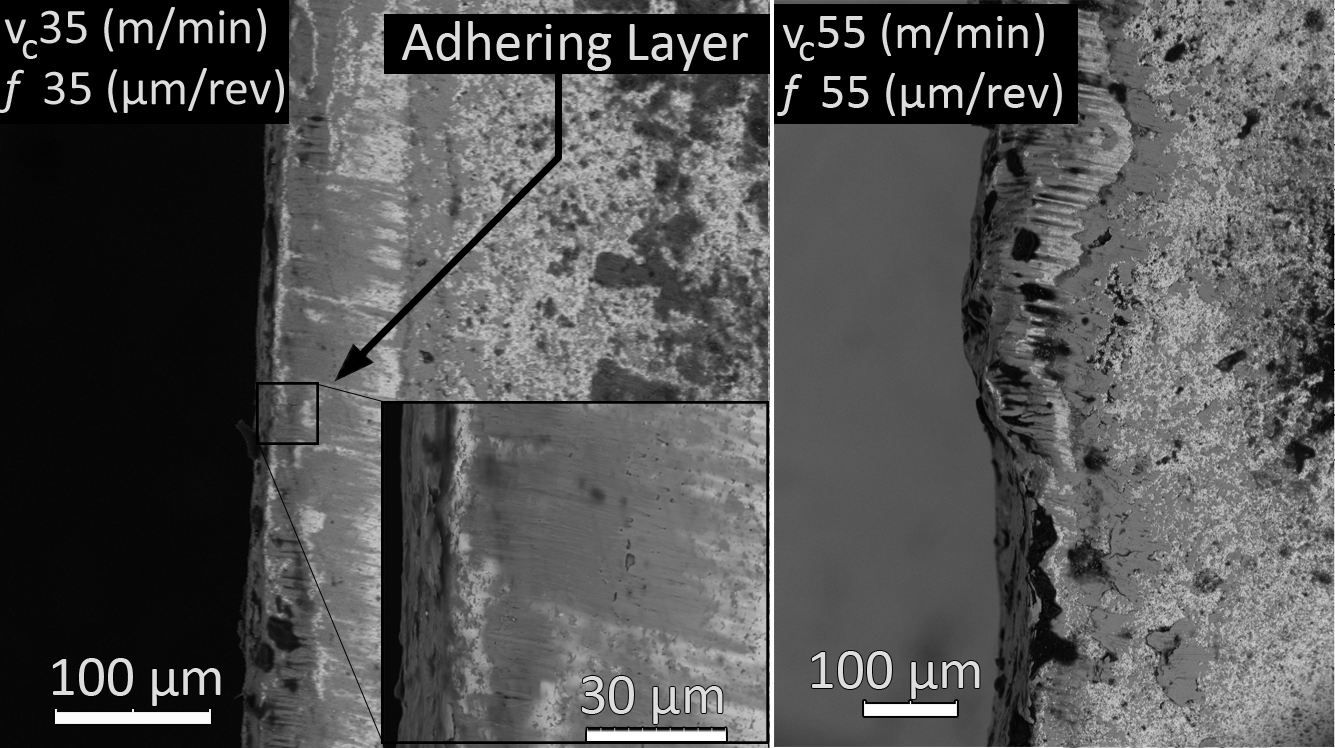}
      \caption{Comparison of worn cutting edge under SEM}
      \label{fig:Wear_Comparison}
    \end{figure}

An examination of the worn cutting tools under scanning electron microscope (SEM) and energy dispersive X-ray spectroscopy (EDX) has indicated the formation of an adhering layer on both the rake and the flank face of the cutting tool and an edge chipping. EDX analysis of this layer has shown a high content of cobalt, chromium and molybdenum. It is most probable that these elements are deposited from the work material in a complex physio-chemical reaction under the extreme conditions pertaining at the tool-workpiece interaction. Moreover, it should be noted that this is an adhering layer and not `built-up edge`. The adhering layer was more extensive when lower cutting speeds were employed and was usually located $<10 \mu m$ from the cutting edge. Higher cutting speeds resulted in chipping of large fragments of the cutting edge. This is shown in Figure~\ref{fig:Wear_Comparison}. 

\begin{figure}[ht]
      \centering
      \includegraphics[width=0.75\textwidth]{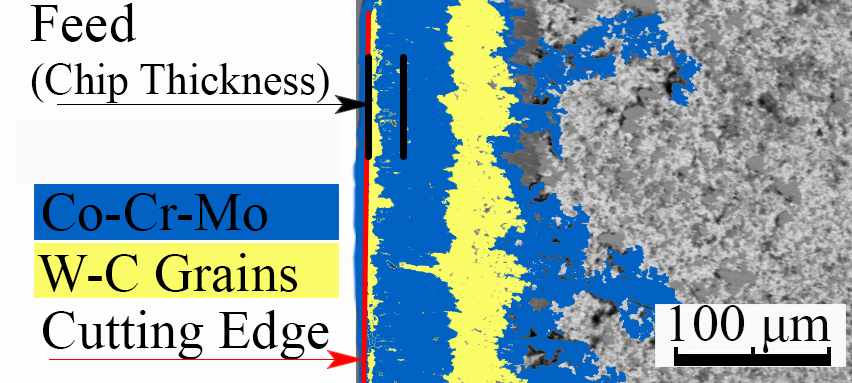}
      \caption{Map of the adhering layers location on the rake face under SEM}
      \label{fig:adhering_layer_map}
    \end{figure}

The characteristic pattern of the adhered layer is shown in Figure~\ref{fig:adhering_layer_map}. The formation of such a layer was also observed when turning Inconel. However, due to the difference in alloy composition, the mechanisms of formation may be different \citep{AdheringLayer}.  As shown in Figure~\ref{fig:adhering_layer_map}, four different zones can be identified. Moving away from the cutting edge in the direction of the chip flow the first zone, a few $\mu m$ wide, is mostly comprised of tungsten and carbon (via EDX analysis). The next zone is comprised  mostly of cobalt, chromium and molybdenum. It is followed by a wider zone containing tungsten and carbon, therefore it is presumed to be the exposed tungsten-carbide. This is followed by a highly irregular layer of adhered cobalt, chromium and molybdenum. This configuration may suggest a sensitivity of this adhering layer to the distribution of the friction/stress/temperature on the rake face of the cutting tool. As the tungsten carbide insert contained a cobalt binder, the mechanism of formation of such layer is presumed due to chemical diffusion/attrition. Specific configuration of the cutting parameters and edge geometry, may form a protective layer and result in flank wear only \citep{itakura1999wear}. In future work, we intend to further explore the mechanisms of tool wear and the performance various tool technologies in cutting of Co-Cr-Mo. Any future results will be compared to the baseline results provided in this paper.

     The tool wear results we have presented thus provide a basis for process control and optimisation for given configurations of the manufacturing process. To further this research, context specific analysis should consider the cost of cutting time versus tool life/cost with a view to optimising the overall manufacturing process chain. Another area of further work will apply the techniques developed here to evaluate the performance of different cutting tools. Optimising tool life in this way could produce significant improvements in overall operating effectiveness and gains in many manufacturing contexts.
  
%

\bibliographystyle{chicago}      
\bibliography{references}   

\end{document}